\documentclass[twocolumn,prc,aps,showpacs]{revtex4}
\usepackage{graphicx}
\usepackage[section]{placeins}

 \def\Pom{{ I\!\!P}}
 
 \def\gsim{\mathrel{\rlap{\lower4pt\hbox{\hskip1pt$\sim$}}
 \raise1pt\hbox{$>$}}}

 \newcommand\beq{\begin{equation}}
 
 \newcommand\eeq{\end{equation}}
 \newcommand\beqn{\begin{eqnarray}}
 \newcommand\eeqn{\end{eqnarray}}
\def\mb{\,\mbox{mb}}

\def\GeV{\,\mbox{GeV}}

\def\lsim{\mathrel{\rlap{\lower4pt\hbox{\hskip1pt$\sim$}}
    \raise1pt\hbox{$<$}}}         
\def\gsim{\mathrel{\rlap{\lower4pt\hbox{\hskip1pt$\sim$}}
    \raise1pt\hbox{$>$}}}         
\def\Re{\,\mbox{Re}\,}
\def\Im{\,\mbox{Im}\,}
\def\mb{\,\mbox{mb}}

\def\GeV{\,\mbox{GeV}}

\def\beq{\begin{equation}}
\def\eeq{\end{equation}}

\def\beqy{\begin{eqnarray}}
\def\eeqy{\end{eqnarray}}

\begin{document}

\title{\bf Single transverse spin asymmetry of forward neutrons}

\author{B. Z. Kopeliovich}
\author{ I. K. Potashnikova}
\author{Iv\'an Schmidt}
\affiliation{\centerline{Departamento de F\'{\i}sica,
Universidad T\'ecnica Federico Santa Mar\'{\i}a; and}
Instituto de estudios avanzados en ciencias en ingenier'a; and\\
Centro Cient\'ifico-Tecnol\'ogico de Valpara\'iso;\\
Casilla 110-V, Valpara\'iso, Chile}
\author{J.~Soffer}
\affiliation{Department of Physics, Temple University, Philadelphia, PA 19122-6082,
USA}

\begin{abstract}
 We calculate the single transverse spin
asymmetry $A_N(t)$, for inclusive neutron production in $pp$
collisions at forward rapidities relative to the polarized proton in
the energy range of RHIC. Absorptive corrections to the pion pole
generate a relative phase between the spin-flip and non-flip
amplitudes, leading to a transverse spin asymmetry which is found to be far too
small to explain the magnitude of $A_N$ observed in the PHENIX
experiment. A larger contribution, which does not vanish at high
energies, comes from the interference of pion and $a_1$-Reggeon
exchanges. The unnatural parity of $a_1$ guarantees a substantial
phase shift, although the magnitude is strongly suppressed by the
smallness of diffractive $\pi p\to a_1 p$ cross section. We replace
the Regge $a_1$ pole by the Regge cut corresponding to the $\pi\rho$
exchange in the $1^+S$ state. The production of such a state, which
we treat as an effective pole $a$, forms a narrow peak in the $3\pi$
invariant mass distribution in diffractive $\pi p$ interactions .
The cross section is large, so one can assume that this state
saturates the spectral function of the axial current and we can
determine its coupling to nucleons via the PCAC Goldberger-Treiman
relation and the second Weinberg sum rule. The numerical results of the
parameter-free calculation of $A_N$ are in excellent agreement with
the PHENIX data.

\end{abstract}



\maketitle

\section{Introduction}\label{intro}

\begin{figure}[!htb]
\centerline{
 {\includegraphics[height=6.5cm]{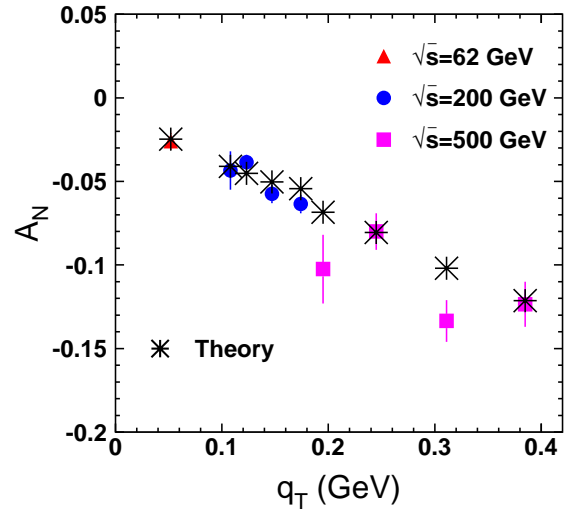}}
 }
\caption{\label{fig:AN-results} (Color online) Single transverse spin asymmetry $A_N$ in the reaction $pp\to nX$,
measured at $\sqrt{s}=62,\ 200,\ 500\GeV$ \cite{phenix} (preliminary
data). The asterisks show the result of our calculation,
Eq.~(\ref{690}), which was done point by point, since each
experimental point has a specific value of $z$ (see
Table~\ref{tab:z-values}).}
 \end{figure}

The single transverse spin asymmetry of neutrons was measured recently by the
PHENIX experiment at RHIC \cite{phenix} in $pp$ collisions at
energies $\sqrt{s}=62,\ 200$ and $500\GeV$. The measurements were
performed with a transversely polarized proton beam and the neutron
was detected at very forward and  backward rapidities relative to
the polarized beam. Preliminary results are depicted in
Fig.~\ref{fig:AN-results}.
 An appreciable single transverse spin asymmetry was found in events with large fractional neutron momenta $z$.
The data agree with a linear dependence on the neutron transverse
momentum $q_T$, and different energy match well, what indicates at
an energy independent $A_N(q_T)$.

Usually polarization data are more sensitive to the mechanisms of reactions than the cross section. Below we demonstrate that the large magnitude of the single transverse spin asymmetry of forward neutrons discovered in \cite{phenix}, reveals a new important mechanism of neutron production ignored in all previous studies of the reaction cross section.

\begin{figure}[!htb]
\centerline{
  \scalebox{0.25}{\includegraphics{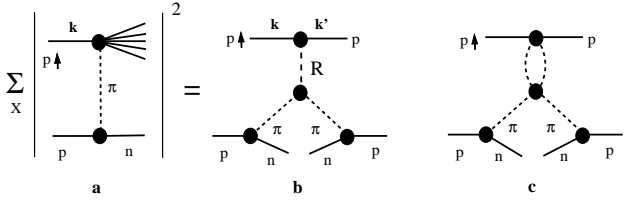}}}
\caption{\label{fig:Ab-Gross}
Graphical representation of the cross
section of inclusive reaction $p\!\!\uparrow\!\!+p\to X+n$ with a polarized proton beam and neutron produced in the fragmentation region of the unpolarized proton, and the related triple Regge graph ({\bf a-b}).
 The Regge cut correction to the previous graph ({\bf c}).}
 \end{figure}

At the same time, neutrons produced with $x_F<0$ show a small
asymmetry, consistent with  zero. This fact is explained by the so
called Abarbanel-Gross theorem \cite{ag} which predicts zero transverse-spin
asymmetry for particles produced in the fragmentation region of an
unpolarized beam. This statement was
proven within the Regge pole model illustrated in Fig.~\ref{fig:Ab-Gross}.
 The amplitude of the reaction $p\!\!\uparrow\!\!+p\to X+n$ squared, Fig.~\ref{fig:Ab-Gross}a,
 is related by the optical theorem with the triple-Regge graph in Fig.~\ref{fig:Ab-Gross}b.
According to Regge factorization
the proton spin can correlates only with the vector product, $[\vec k\times\vec k^{\prime}]$, of the proton momenta in the two conjugated amplitudes, as is shown in Fig.~\ref{fig:Ab-Gross}b.
According to the optical theorem these momenta are equal, $\vec k=\vec k^\prime$,
so no transverse spin correlation is possible. 
Regge cuts shown in Fig.~\ref{fig:Ab-Gross}c breakdown this statement, but the magnitude of the gained single-spin asymmetry calculated in \cite{klz}, turned out to be very small, less than $1\%$.

\section{Spin structure of the pion pole}

\begin{figure}[tbh]
\centerline{
  \scalebox{0.41}{\includegraphics{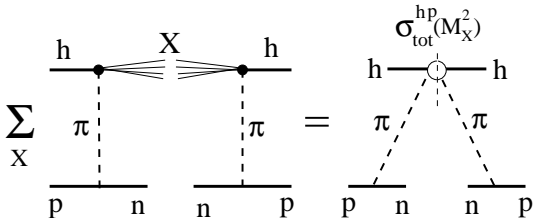}}}
\caption{\label{fig:3r-pion} Graphical representation of the cross
section of inclusive neutron production in hadron-proton collisions,
in the fragmentation region of the proton. }
 \end{figure}

The importance of the pion pole for neutron production has been
known since the pioneering paper \cite{bishari}. Its Reggeized
version for leading neutrons in DIS was proposed in \cite{k2p}. A
further development, which includes the spin structure of the
amplitudes was published in \cite{kpss1}. Contrary to the
conventional wisdom that the pion-nucleon vertex is pure spin-flip,
a large non-flip term in the amplitude appears if $z<1$. In Born
approximation the pion exchange in neutron production, depicted in
Fig.~\ref{fig:3r-pion}, in the leading order in the small parameter
$m_N/\sqrt{s}$ has the form \cite{kpss1},
 \beq
A^B_{p\to n}(\vec q,z)=
\bar\xi_n\left[\sigma_3\, q_L+
\frac{1}{\sqrt{z}}\,
\vec\sigma\cdot\vec q_T\right]\xi_p\,
\phi^B(q_T,z)\,,
\label{100}
 \eeq
 where $\vec\sigma$ are the Pauli matrices;  $\xi_{p,n}$ are the proton or
neutron spinors;  $\vec q_T$ and
 \beq
q_L=\frac{1-z}{\sqrt{z}}\,m_N\,,
\label{110}
 \eeq
are  the transverse and longitudinal components of the momentum transfer respectively.

At large $z$ the pseudoscalar amplitude $\phi^B(q_T,z)$ has
the Regge form \cite{k2p},
 \beqn
\phi^B(q_T,z)&=&\frac{\alpha_\pi^\prime}{8}\,
G_{\pi^+pn}(t)\,\eta_\pi(t)\,
(1-z)^{-\alpha_\pi(t)}
\nonumber\\ &\times&
A_{\pi^+ p\to X}(M_X^2)\,,
\label{120}
 \eeqn
 where $M_X^2= (1-z)s$; the 4-momentum
transfer squared ($t$) has the form,
 \beq
-t=q_L^2+{1\over z}\,q_T^2;
\label{130}
 \eeq
 and $\eta_\pi(t)$ is the phase (signature) factor,  \beq
\eta_\pi(t)=i-ctg\left[\frac{\pi\alpha_\pi(t)}{2}\right]\,.
\label{140}
 \eeq
 The second term contains the pion pole,
 \beq
 \Re\eta_\pi(t)\approx
 \frac{2}{\pi\alpha_\pi^\prime}\,
\frac{1}{m_\pi^2-t}.
\label{140n}
\eeq
 In what follows we assume that the pion Regge trajectory is a linear function of $t$,
$\alpha_\pi(t)=\alpha_\pi^\prime(t-m_\pi^2)$,
where $\alpha_\pi^\prime\approx 0.9\GeV^{-2}$.

The effective vertex function is parametrized as, \beq
G_{\pi^+pn}(t)=g_{\pi^+pn}\,e^{R_\pi^2t}, \label{140nn} \eeq where
the pion-nucleon coupling $g^2_{\pi^+pn}/8\pi=13.85$. The $t$-slope
parameter $R_\pi^2$ incorporates the $t$-dependences of the coupling
and of the $\pi N$ inelastic amplitude. Although it is not well
known, its value is not really important for us, since we
concentrate on the small $t$  region. For further calculations we
fix $R_\pi^2=4\GeV^{-2}$, which is naturally related to the nucleon
size.

The amplitude (\ref{100}) is normalized as $M_X^2\sigma^{\pi^+
p}_{tot}= \sum\limits_X|A_{\pi^+ p\to X}(M_X^2)|^2.$
Correspondingly, the differential cross section of inclusive neutron
production reads,
 \beqn
z\,\frac{d\sigma^B_{p\to n}}{dz\,dq_T^2}&=&
\left(\frac{\alpha_\pi^\prime}{8}\right)^2
|t|G_{\pi^+pn}^2(t)\left|\eta_\pi(t)\right|^2
\nonumber\\ &\times&
(1-z)^{1-2\alpha_\pi(t)}
\sigma^{\pi^+ p}_{tot}(M_X^2)\,.
\label{146}
 \eeqn

 This pure pion pole model has two obvious shortcomings: (i) the cross section Eq.~(\ref{146}) substantially overshoots data \cite{kpss1}; (ii) no single transverse spin asymmetry is possible because the spin-flip and non-flip terms in the amplitude (\ref{100}) have no phase shift.

 \section{Spin effects generated by initial/final state interactions}

It is known that 
initial and final state interactions (ISI/FSI) can generate spin effects \cite{ivan-stan}, if they are absent in the Born approximation. In the process under discussion ISI/FSI
essentially modify the spin amplitudes of the process, suppressing the magnitude of the cross section and generating a nonzero relative phase shift.

Absorptive corrections look simpler and factorize in impact
parameter. Therefore, we perform a Fourier transform on the amplitude (\ref{100})
to impact parameter representation, multiply it by the survival
probability amplitude $S(b,z)$,
 \beq
f_{p\to n}(b,z)=f^B_{p\to n}(b,z)\,S(b,z),
\label{195}
 \eeq
 and then return back to the momentum representation \cite{kpss1}. The Born amplitude (\ref{100}) in impact parameter reads,
 \beq
f^B_{p\to n}(\vec b,z)=
\bar\xi_n\left[\sigma_3\, q_L\,\theta^B_0(b,z)-
i\,\frac{\vec\sigma\cdot\vec b}{\sqrt{z}\,b}\,
\theta^B_s(b,z)\right]\xi_p,
\label{150}
 \eeq
 where
\begin{widetext}
\beqn
\theta^B_0(b,z) &=& \int d^2q\,e^{i\vec b\vec q}\,
\phi^B(q_T,z)= N^\pi_X(z) 
\Biggl\{i\,\frac{\pi\alpha_\pi^\prime}{2z\beta^2}\,
K_0(b/\beta) +
\frac{1}{1-\beta^2\epsilon^2}\,
\left[K_0(\epsilon b)-K_0(b/\beta)\right]\Biggr\}\,;
\label{154}\\
\theta^B_s(b,z) &=& {1\over b}
\int d^2q\,e^{i\vec b\vec q}\,
(\vec b\cdot\vec q)\,\phi^B(q_T,z)=
N^\pi_X(z)
\Biggl\{i\,\frac{\pi\alpha_\pi^\prime}{2z\beta^3}\,
K_1(b/\beta)
+\frac{1}{1-\beta^2\epsilon^2}\,
\left[\epsilon\,K_1(\epsilon b)-\frac{1}{\beta}\,K_1(b/\beta)
\right]\Biggr\}\,;
\label{164}
 \eeqn
\end{widetext}

and
 \beq
N^\pi_X(z) =\frac{1}{2}\,g_{\pi^+pn}\,
z(1-z)^{\alpha^\prime_\pi(m_\pi^2+q_L^2)}
e^{-R_\pi^2 q_L^2}
A_{\pi p\to X}(M_X^2)
\label{166}
\eeq
\beqn
\epsilon^2&=& z(q_L^2+m_\pi^2)\,,
\nonumber\\
\beta^2&=&{1\over z}\,\left[
R_\pi^2-\alpha_\pi^\prime\,\ln(1-z)\right]\,.
\label{166a}
 \eeqn

As is illustrated in Fig.~\ref{fig:3r-pion}, in the rapidity
interval covered by the pion, $\Delta y=|\ln(1-z)|$, no particles
are produced. Since inelastic ISI/FSI of
the spectator partons tend to fill this rapidity gap, its survival
probability leads to a suppression. It was found in \cite{kpss1}
that the partonic system, which participates in ISI/FSI with the
target nucleon, is a 5-quark color octet-octet dipole. The amplitude
suppression factor $S(b,z)$ was found in \cite{kpss1} to be rather
small, leading to a strong reduction of the cross section of leading
neutron production. The magnitude of the partial cross section,
which is expected to be universal, was found to agree well with ZEUS
data for neutron produced in DIS \cite{zeus}, and with NA49 data for
$pp$ collisions \cite{na49}. Although these data confirm the shape
of the $z$ distribution measured earlier at ISR \cite{isr}, the
latter overestimated the normalization.

Following (\ref{195}) we applied the absorption factor to the Born spin amplitudes Eq.~({\ref{150}),
and performing Fourier transfer back to momentum representation, arrive at,
 \beq
A_{p\to n}(\vec q,z)=
\bar\xi_n\left[\sigma_3 q_L\,\phi_0(q_T,z)-
i\vec\sigma\cdot\vec q_T\frac{\phi_s(q_T,z)}{\sqrt{z}}\right]\xi_p,
\label{520}
 \eeq
 where
   \beqn
\Re\phi_0(q_T,z)&=&\frac{N^\pi_X(z)}{2\pi(1-\beta^2\epsilon^2)}
\int\limits_0^\infty db\,b\,J_0(bq_T)
\nonumber\\ &\times&
\left[K_0(\epsilon b)-K_0(b/\beta)\right]
S(b,z);
\label{550a}\\
\Im\phi_0(q_T,z)&=&\frac{\alpha_\pi^\prime N^\pi_X(z)}{4z\beta^2}
\int\limits_0^\infty db\,b\,J_0(bq_T)\,
\nonumber\\ &\times&
K_0(b/\beta)\,S(b,z);
\label{550b}
 \\
 \Re\phi_s(q_T,z)&=&\frac{N^\pi_X(z)}{2\pi(1-\beta^2\epsilon^2)q_T}
\int\limits_0^\infty db\,b\,J_1(bq_T)
\nonumber\\ &\times&
\left[\epsilon K_1(\epsilon b)-
{1\over\beta}K_1(b/\beta)\right]S(b,z);
\label{570a}\\
\Im\phi_s(q_T,z)&=&\frac{\alpha_\pi^\prime N^\pi_X(z)}{4z\beta^3q_T}
\int\limits_0^\infty db\,b\,J_1(bq_T)\,
\nonumber\\ &\times&
K_1(b/\beta)\,
S(b,z).
\label{570b}
 \eeqn

We are now in a position to calculate the single transverse spin asymmetry,
 \beqn
A_N(q_T,z)&=&\frac{2q_T q_L\sqrt{z}
\sum_X
\left|\phi_0(q_T,z)\right|
\left|\phi_s(q_T,z)\right|}
{z q_L^2 \sum_X\left|\phi_0(q_T,z)\right|^2 +
q_T^2 \sum_X\left|\phi_s(q_T,z)\right|^2}
\nonumber\\ &\times&
\sin(\delta_s-\delta_0)\,,
\label{620}
 \eeqn
 where
 \beq
\tan\delta_{0,s}=\frac{\Im\phi_{0,s}(q_T,z)}
{\Re\phi_{0,s}(q_T,z)}\,.
\label{640}
 \eeq
The results are depicted in Fig.~\ref{fig:AN-pion}.
 \begin{figure}[tbh]
\centerline{
{\includegraphics[height=6cm]{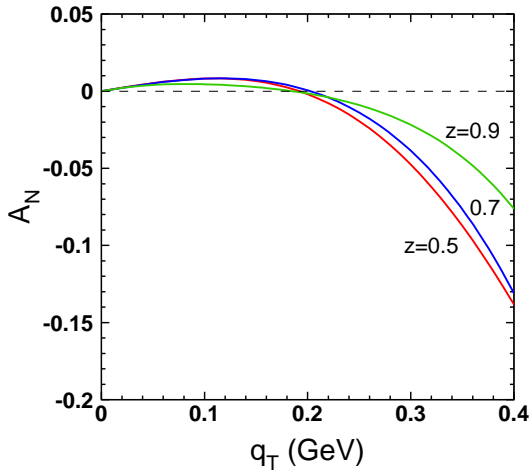}}
}
 \caption{\label{fig:AN-pion} (Color online)
 Single transverse spin asymmetry of leading neutrons related to single pion exchange corrected for absorptive corrections, as function of $q_T$. The curves from bottom to top correspond to $z=0.5,\ 0.7$ and $0.9$.}
 \end{figure}
It is clear that  our calculations significantly underestimate the PHENIX data in the range of
$q_T$ covered by the experiment.

 \section{Other sources of spin effects: 
 \newline \boldmath$a_1$ Reggeon?}

 In addition to pion exchange, other Regge poles $R=\rho,\ a_2,\ \omega,\ \ a_1$, etc.
 and Regge cuts can contribute to the $pp\to nX$ reaction.
 We should remind that the total collision energy is shared by the reaction
$\pi+p\to R+p$ depicted in the upper blob of the diagram
Fig.~(\ref{fig:pi-R}), and by the rapidity gap, rather unequally:
while the former can be very large, $M_X^2=(1-z)s$ at high energies,
the c.m. energy for pion exchange is rather small,
$s^\prime=s_0/(1-z)$, and independent of $s$.

  \begin{figure}[!thb]
\centerline{
  \scalebox{0.3}{\includegraphics{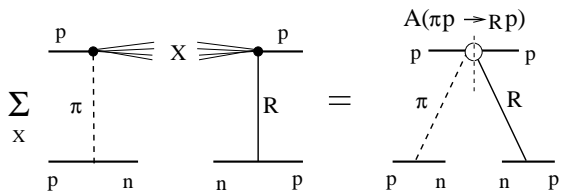}}
}
 \caption{\label{fig:pi-R} Graphical representation for the interference between the amplitudes with pion and Reggeon exchanges.
}
 \end{figure}
Summing over different produced states $X$ and using completeness
one arrives at the imaginary part of the amplitude of the process
$\pi+p\to R+p$ at c.m. energy $M_X^2$. The production of natural
parity states, like $\rho$, $a_2$, etc. can proceed only via reggeon
exchange, therefore these amplitudes are strongly suppressed at RHIC
energies by a power of $M_X$ (dependent on the Regge intercept) and
can be safely neglected everywhere, except at the region of very
small $(1-z)\sim s_0/s$, unreachable experimentally.

 Only the unnatural parity states, which can be diffractively produced by a pion,
like the $a_1$ meson, or $\rho$-$\pi$ in the axial vector or
pseudo-scalar states, contribute to the interference term in the
neutron production cross section at high energies.

The $a_1NN$ vertex is known to be pure non spin-flip \cite{owens,kane}. Therefore, it should be added to the first term in Eq.~(\ref{520}),
 \beq
A_{p\to n}^{a_1}(q_T,z)=
e_\mu^L\,\bar n\,\gamma_5\gamma_\mu\,p =\frac{2m_Nq_L}{\sqrt{|t|}}
\,\phi_0^{a}(q_T,z)\,\bar\xi_n\sigma_3\xi_p,
\label{660}
 \eeq
where the longitudinal polarization vector of $a_1$ reads \cite{marage},
\beq
e_\mu^L=\frac{1}{\sqrt{|t|}}\,\left(\sqrt{q_0^2-t},\,0,\,0,\,q_0\right),
\label{666}
\eeq
and the transferred energy
\beq
q_0=E_p-E_n=q_L + O\!\left(m_N/\sqrt{s}\right).
\label{668}
\eeq
In the Born approximation,
 \beqn
\phi^a_0(q_T,z)&=&\frac{\alpha_{a_1}^\prime}{8}\,
G_{a^+pn}(t)\,\eta_{a_1}(t)
\nonumber\\ &\times&
(1-z)^{-\alpha_{a_1}(t)}
A_{a_1^+ p\to X}(M_X^2)\,,
\label{670}
 \eeqn
and
\beq
\eta_{a_1}(t)=-i-tg\!\left[\frac{\pi\alpha_{a_1}(t)}{2}\right].
\label{680}
\eeq

The amplitude (\ref{660}) contains three unknowns, to be fixed
before numerical evaluation:
\begin{itemize}
\item
The amplitude $A_{a_1^+ p\to X}(M_X^2)$;
\item
The $a_1$-nucleon vertex $G_{a_1^+pn}(t)$;
\item The Regge trajectory $\alpha_{a_1}(t)$.
\end{itemize}

Notice that the general structure of the amplitude Eq.~(\ref{670}) is valid for any axial-vector state. In what follows we find the $a_1$ pole to be quite a weak singularity, and conclude that the spectral function of the axial current is dominated by the contribution of $\pi\rho$ pair in the $1^+S$ state. So we will replace the weak $a_1$ pole by an effective singularity $a$.

\subsection{Diffractive production of \boldmath$a_1$ meson and non-resonant 
\boldmath$\pi\rho(1^+S)$ pairs}

The amplitude $A_{a_1^+ p\to X}(M_X^2)$ is normalized as,
\beqn
&&\sum\limits_X A^\dagger_{a_1^+ p\to X}(M_X^2)\,
A_{\pi p\to X}(M_X^2)=4\sqrt{\pi}\,M_X^2
\nonumber\\ &\times&
\sqrt{\left.d\sigma(\pi p\to a_1p)/dp_T^2\right|_{p_T=0}}
\label{685b}
\eeqn

Evaluation of the ratio of the forward diffractive-to-elastic cross
sections in this expression is a subtle problem. The cross section
of diffractive $a_1$ production is so small that it has been
escaping detection for a long time. Eventually $a_1$ production was
observed in $\pi+p\to3\pi+p$ performing a phase shift analysis of
high-statistics data \cite{antipov,1979}. The $a_1$ resonance was
only detected in this process as a fast variation of the relative
phase of $1^+S$.

 The $\rho\pi$$(1^+S)$ mass distribution depicted in Fig.~\ref{fig:3pi} 
forms a strong and narrow peak, related mainly to the Deck mechanism
\cite{deck}.
\begin{figure}[!htb]
\centerline{
 {\includegraphics[height=6.5cm]{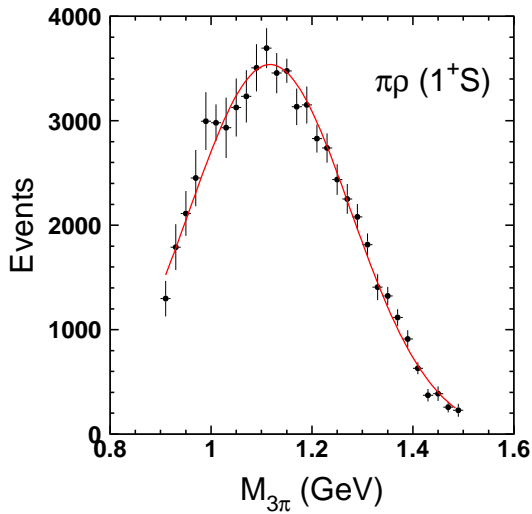}}
 }
\caption{\label{fig:3pi} (Color online)
Variation of $1^+S$ wave intensity with $3\pi$ invariant mass, for
$q_T^2<0.05\GeV^2$. Data for
$E_{lab}= 63$ and $94\GeV$ are summed \cite{1979}.
The curve is fitted with parametrization (\ref{fit}).}
 \end{figure}
The data \cite{1979} combined for lab energies $63$ and $94\GeV$ were taken with the small-$q_T$ trigger. We normalize this distribution on the measured cross section
in the invariant mass interval $1.0<M_{3\pi}<1.2\GeV$ which is $54\pm8\,\mu b$ and $48\pm7\,\mu b$ at $63$ and $94\GeV$ respectively. With the measured $p_T^2$-slope 
$B=14.3\pm0.17\GeV^{-2}$ we can evaluate the forward cross section for this mass interval at $d\sigma/dp_T^2\bigr|_{p_T=0}=0.73\pm0.07\mb/\!\GeV^2$ averaged for the two energies.

In order to normalize the mass distribution plotted in  Fig.~\ref{fig:3pi} to this cross section, we fitted the data with a simple Gaussian parametrization,
\beq
\frac{d\sigma}{dM_{3\pi}\,dq_T^2}\Bigr|_{q_T=0}=
\lambda\exp\left[-\frac{(M_{3\pi}-m_a)^2}{\Gamma_a^2}\right],
\label{fit}
\eeq
with parameters $m_a=(1.117\pm 0.004)\GeV$, $\Gamma_a=(0.226\pm 0.004)\GeV$, and
$\lambda=3.91\pm 0.03\mb/\GeV^3$. We also made a $10\%$ correction for the nonzero values of $q_T^2<0.05\GeV^2$ of data in Fig.~\ref{fig:3pi}.
Notice that the central mass of the peak is  significantly smaller than $m_{a_1}=1.28\GeV$ \cite{pdg}. 

Eventually, integrating the normalized distribution over the invariant $3\pi$ mass under the
$a$-peak in Fig.~\ref{fig:3pi} we arrive at the $\pi\rho(1^+S)$
forward production cross section, 
\beq 
\frac{d\sigma(\pi p\to\pi\rho p)}
{dp_T^2}\Bigr|_{1^+S,p_T=0}=1.67\mb/\!\GeV^2.
\label{a10} 
\eeq

Now we have to extrapolate the diffractive cross section
Eq.~(\ref{a10}), from the energy range of the fixed target
experiments \cite{1979}, $s_1\sim 119-177\GeV^2$ up to the
substantially higher c.m. energies $M_X^2=(1-z)s$ of RHIC. We
extrapolate the single diffractive cross section with the following
energy dependence, \beq \frac{d\sigma_{sd}(M_X^2)}{dq_T^2}=
\frac{d\sigma_{sd}(s_1)}{dq_T^2}\,\left(\frac{M_X^2}{s_1}\right)^{0.16}
\,\frac{K_{\pi p}(M_X^2)}{K_{\pi p}(s_1)}, \label{a20} \eeq where
$K(s)$ is the survival probability of a large rapidity gap, which in
the eikonal approximation has the form \cite{DY-diff}
 \beqn
K_{\pi p}(s)&=&1-{1\over{\pi}}\,\frac{\sigma^{\pi p}_{tot}(s)}
{B^{\pi p}_{sd}(s)+2B^{\pi p}_{el}(s)}
\nonumber\\ &+&
 \frac{1}{(4\pi)^2}\,
\frac{\left[\sigma^{\pi p}_{tot}(s)\right]^2}
{B^{\pi p}_{el}(s) \left[B^{\pi p}_{sd}(s)+B^{\pi p}_{el}(s)
\right]}.
 \label{a30}
 \eeqn

For the energy dependent total $\pi p$ cross section $\sigma^{\pi
p}_{tot}(s)$ we rely on the parametrization given in \cite{pdg}. For
the elastic and single diffractive slopes we use $B^{\pi
p}(s)=B_0+2\alpha_{\Pom}^\prime\,\ln(s/s_0)$, where $B_0=6\GeV^{-2}$
for elastic and $B_0=9\GeV^{-2}$ for single diffractive processes.
$\alpha_{\Pom}^\prime=0.25\GeV^{-2}$.

Notice that the replacement of the $a_1$ pole in the dispersion
relation by the effective singularity $a$ also helps to settle
several problems related to the assumed $a_1$ dominance for the
axial current. In particular, this assumption leads to so called
Piketty-Stodolsky paradox \cite{piketty,marage}. Namely, based on
the PCAC hypothesis, which relates the pion pole and heavier states
contributions in the dispersion relation for the axial current, one
arrives at the conclusion that diffractive $\pi\to a_1$ and elastic
$\pi\to\pi$ cross sections should be equal, while experimentally
they are different by more than one order of magnitude. Replacing
the $a_1$ pole by the effective singularity $a$ solves the problem
\cite{belkov,marage,kpss-nu,kps-nu}.

Summarizing, we found that the production cross section for the
$a_1$ meson is too small to produce a sizable contribution to
neutron production. It should be replace by the more significant
production of a $\pi\rho$ $1^+S$ state, which forms a narrow
resonance-like peak in the $3\pi$ invariant mass distribution. Thus,
we introduce an effective "pole" $a$ in the dispersion relation for
the axial-vector current, and predict its production cross section
in $\pi p$ collisions at high energies.

\subsection{\boldmath$aNN$ coupling}

Like for the pion we parametrize the $a_1$-nucleon vertex in
Eq.~(\ref{670}) as $G_{a^+pn}(t)=g_{a^+pn}\exp(R_a^2\,t)$. The slope
parameter $R_a^2$ is poorly known, however in the small $t$ region
under consideration it is not of great importance. Like for the pion
vertex, it is natural to expect the slope to be related to the
nucleon size, so we fix $R_a^2=R_\pi^2=4\GeV^2$ for further
calculations.

The $a$-nucleon coupling $g_{a^+pn}$ can be estimated based on PCAC.
Although it is tempting to interpret the Goldberger-Treiman relation
and Adler theorem as pion pole dominance, the pion pole does not
contribute in either $\beta$-decay or high-energy neutrino
interactions, because of conservation of the lepton current
(neglecting the lepton mass) \cite{bell,piketty,marage}. In order to
have PCAC  heavy axial states contributing to the dispersion
relation for the amplitude of the process must miraculously
reproduce the pion pole. If we replace the combined contribution of
the heavy state by an affective pole $a$
\cite{belkov,marage,kpss-nu}, the Goldberger-Treiman relation
relates the $a$ and pion poles, \beq
\frac{\sqrt{2}f_a\,g_{aNN}}{m_a^2}=\frac{f_\pi\,g_{\pi
NN}}{\sqrt{2}m_N} \label{686} \eeq

In the second Weinberg sum rule the spectral function of the vector
current can be represented by the $\rho$-meson pole.
Correspondingly, the axial spectral function is saturated by the $a$
meson, because the pion does not contribute to the second Weinberg
sum rule. Then one arrives at the relation, \beq
f_a=f_\rho=\frac{\sqrt{2}m_\rho^2}{\gamma_\rho}, \label{687} \eeq
where $\gamma_\rho$ is the universal coupling ($\rho NN$,
$\rho\pi\pi$, etc), $\gamma_\rho^2/4\pi=2.4$.

Thus, for the $a$ to pion couplings ratio we get,
\beq
\frac{g_{a NN}}{g_{\pi NN}}=
\frac{m_a^2\,f_\pi}{2m_N\,f_\rho}\approx 0.5.
\label{686a}
\eeq

A different expression was derived in \cite{kane}, based on the
first Weinberg sum rule, which includes the pion pole,
\beq
\frac{g_{aNN}}{g_{\pi NN}}= \frac{m_{a}m_\rho f_\pi}
{2m_N(f_\rho^2-m_\rho^2f_\pi^2)^{1/2}}. \label{688} \eeq
Numerically
the results (\ref{686a}) and (\ref{688}) are very close in the
chiral limit of massless pion. Otherwise they differ by $15\%$.

\subsection{Regge trajectory of the "\boldmath$a$-pole"}

So far  the narrow $a$-peak in the spectral function of the axial
current could be treated as an effective pole replacing the real one
$a_1$, which was found to be too weak. However,  the Regge
singularity in the complex angular momentum plane, related to the
$\pi$-$\rho$ exchange, is a Regge cut rather than a pole. The
trajectory of the cut can be expressed in terms of the $\pi$ and
$\rho$ Reggeons, \beq
\alpha_{\pi\rho}(t)=\alpha_\pi(0)+\alpha_\rho(0)-1+
\frac{\alpha_\pi^\prime\alpha_\rho^\prime}
{\alpha_\pi^\prime+\alpha_\rho^\prime}\,t. \label{689} \eeq For
further numerical evaluations we fix $\alpha(0)=1/2$ and
$\alpha_\pi^\prime=\alpha_\rho^\prime=0.9\GeV^{-2}$, so
$\alpha_{\pi\rho}(t)=-0.5+0.45t$.

Correspondingly, the phase (signature) factor for the unnatural parity $a$-Reggeon exchange reads,
\beq
\eta_{a}(t)=-i-tg\left[\pi\alpha_a(t)/2\right],
\label{189a}
\eeq
where $\alpha_{a}(t)=\alpha_{\pi\rho}(t)$.

This factor provides a significant phase shift  $\Delta\phi=\pi/4$
relative to the pion, and this phase shifts rises with $t$ up to the
maximal value of $\pi/2$ at $t=-1\GeV^2$. This interference looks
like a promising source of a significant single transverse spin asymmetry.

Notice that the intercept of the $\pi$-$\rho$ cut turns out to be rather close to the intercept of the $a_1$ Regge pole, $\alpha_{a_1}(0)=-0.43$, which corresponds to a straight  Regge trajectory with the universal slope crossing the position of the $a_1$ pole on the Chew-Frautschi plot.\\

\section{Single transverse spin asymmetry generated by \boldmath$\pi$-$a$ interference}

Eventually, we are in a position to perform a parameter free calculation of the  $a$-$\pi$ interference contribution to single transverse spin asymmetry of neutron production,
 \beqn
&&A_N(q_T,z) =
q_T\frac{4m_N\,q_L}{|t|^{3/2}}
(1-z)^{\Delta\alpha(t)}
\frac{\Im\eta_\pi^*(t)\,\eta_{a}(t)}
{\left|\eta_\pi(t)\right|^2}\,
\nonumber\\ &\times&
\frac{g_{a^+pn}}{g_{\pi^+pn}}
\left(\frac{d\sigma_{\pi p\to a p}(M_X^2)/dp_T^2|_{p_T=0}}
{d\sigma_{\pi p\to\pi p}(M_X^2)/dp_T^2|_{p_T=0}}\right)^{1/2},
\label{690}
 \eeqn
 where $\Delta\alpha(t)=\alpha_\pi(t)-\alpha_a(t)$.

The specific value of $z$ is important for the factors that appear
in (\ref{690}), in particular for $t$ given by
(\ref{110})-(\ref{130}). We extracted values of $z$ for each
experimental point shown in Fig.~\ref{fig:AN-results}) from data
\cite{phenix} as is described in \ref{app:z} and  presented in
Table~\ref{tab:z-values}. The results of the calculation with
Eq.~(\ref{690}) are plotted in Fig.~\ref{fig:AN-results} by
asterisks. The agreement with data is excellent, but we estimate the
uncertainty of the normalization of the theoretical results at about
$30\%$.

Notice that in these calculations we neglected the absorptive
corrections, because they have very little influence on $A_N$,
Eq.~(\ref{690}). Indeed, as we have seen above, they cause a tiny
variation of the relative phase, which can be neglected. It was also
demonstrated in \cite{kpss1} that the spin-flip and non-flip
amplitudes are suppressed by absorption similarly, because the
kinematics of the data is rather far from the pion pole (see
Table~\ref{tab:z-values}). We checked that inclusion of absorption
leaves the asymmetry well within the theoretical uncertainty of the
calculations.

Although the PHENIX experiment also provided data for $A_N$ 
with a special selection of charged hadrons detected at pseudo-rapidities
$3.0<\eta<3.9$, we do not think that comparison with these data would bring
new valuable information. Indeed, in this case summing over the intermediates states $X$
one cannot  rely on completeness any more, and the theoretical uncertainties
would increase significantly. 

 \section{Summary}

The results of the the PHENIX experiment at RHIC on single transverse spin
asymmetry of neutrons produced in polarized $pp$ collisions at very
forward rapidities have brought precious information about the
mechanisms of leading neutron production. The naturally expected
pion pole calculated in the Born approximation does not explain
either the magnitude of the cross section, nor the observed
azimuthal asymmetry. The absorptive corrections, which can be
interpreted as the survival probability of a rapidity gap,
substantially reduce the cross section improving agreement with
data. At the same time, the relative phase shift between the
spin-flip and non-flip amplitudes generated by absorption turns out
to be far too small to explain the measured single transverse spin asymmetry.

Another possible source of a considerable spin effect
considered here is the interference between the amplitude of neutron
production via pion and $a_1$ Reggeon exchanges. Because $a_1$  has
unnatural parity, it can be produced diffractively in $\pi+p\to
a_1+p$, so is not suppressed at high c.m. energy $M_X$. It also
provides a large, close to maximal, relative phase shift between the
non-flip $a_1$ and spin-flip pion exchange amplitudes.

It turns out, however, that the $a_1$ exchange contribution is
strongly suppressed by the smallness of the diffractive $a_1$
resonance production. Nevertheless, we found that it is possible to
replace this resonance by $\pi\rho$ in the unnatural parity $1^+S$
state, because it forms a narrow peak in the $3\pi$ invariant mass
distribution, so can be treated as an effective pole, named $a$, in
a dispersion relation for the axial current.

This allows to determine the $a$-nucleon coupling using PCAC, which
relates the contributions of heavy states (saturated by the $a$ pole
and the pion pole). Additional information about the leptonic decay
constant of $a$ is obtained from the second Weinberg sum rule.

Although the $\pi\rho$ exchange corresponds to a Regge cut, rather than a pole, we found its Regge intercept to be rather close to the one for $a_1$ Reggeon, so the phase shift is similar as well.

Finally, we calculated the single transverse spin asymmetry at different values
of the kinematic variables, $s$, $q_T$ and $z$, and found very good
agreement with data.

\section{Acknowledgments}

We are thankful to Elke Aschenauer and Yousef Makdisi for providing information about the
new PHENIX results on neutron single transverse spin asymmetry.
This work was supported in part by Fondecyt (Chile) grants 1090236,
1090291 and 1100287,  and by Conicyt-DFG grant No. 084-2009.\\


 \def\appendix{\par
 \setcounter{section}{0}
\setcounter{subsection}{0}
 \def\thesection{Appendix \Alph{section}}
\def\thesubsection{\Alph{section}.\arabic{subsection}}
\def\theequation{\Alph{section}.\arabic{equation}}
\setcounter{equation}{0}}

 \appendix

\section{Fractional neutron momenta in PHENIX data}
\label{app:z} \setcounter{equation}{0}

We extracted values of $z$ for each experimental point shown in
Fig.~\ref{fig:AN-results}, from data \cite{phenix}, using the
relation with the scattering angle, $q_T=z\sin\theta\sqrt{s}/2$. The
results for all points are presented in Table~\ref{tab:z-values}.

\begin{table}[!ht]
 \begin{center}
\caption{\label{tab:z-values} Values of angles of neutron detection,
of neutron transverse momenta, and deduced values of fractional
neutron momenta $z$, in the results of the PHENIX measurements
\cite{phenix} of the neutron single transverse spin asymmetry.}
\end{center}
\begin{tabular*}
{0.48\textwidth}{@{\extracolsep{\fill}}|c| c| c c c c| c c c c|}\hline
$\sqrt{s}$(GeV)& $62$ &  & $200$ & & & & $500$ & & \\ \hline
$\theta$(mrad)& 1.89& 1.38& 1.57& 1.87& 2.21& 1.083& 1.36& 1.74& 2.15
\\ \hline
$q_T$(GeV) &0.05  & 0.11 & 0.12 & 0.15 & 0.18& 0.20& 0.25& 0.31& 0.39 \\
\hline $z$  & 0.89  & 0.78 & 0.78 & 0.79 & 0.79& 0.72& 0.73& 0.72& 0.72\\
\hline 
 \end{tabular*}
\end{table}


\bibliographystyle{aipproc}

\end{document}